\documentclass[%
superscriptaddress,
preprint,
 amsmath,amssymb,
 aps, physrev,
]{revtex4-2}

\usepackage{blindtext}
\usepackage{float}
\usepackage[acronym, shortcuts, nowarn]{glossaries}
\usepackage{graphicx}
\usepackage{dcolumn}
\usepackage[version=4]{mhchem}
\usepackage[separate-uncertainty=true]{siunitx}
\usepackage{xcolor}
\usepackage{cleveref}
\usepackage{amsmath}
\usepackage{kpfonts}
\usepackage{subfiles}
\newcommand{\AtiAddress}{Advanced Technology Institute, University of Surrey, Guildford, GU2 7XH, UK}
\newcommand{\CardiffAddress}{School of Physics \& Astronomy, Cardiff University, Queen's Buildings, Cardiff, CF24 3AA, UK}
\newcommand{\HFMLAddress}{High Field Magnet Laboratory (HFML-EMFL), Radboud University, Toernooiveld 7, 6525 ED Nijmegen, Netherlands}
\newcommand{\IMM}{Institute for Molecules and Materials, Radboud University, Heyendaalseweg 135, 6525 AJ Nijmegen, The Netherlands.}

\makeglossaries
\setacronymstyle{long-short}

\begin{document}

\title{Absolute measurement of the exchange interaction in an InSb quantum well using Landau-level tunnelling spectroscopy}
\author{S.K. Clowes}    \affiliation{\AtiAddress}
\author{C.P Allford} \affiliation{\CardiffAddress}
\author{D. Shearer} \affiliation{\AtiAddress}
\author{G.V. Smith}       \affiliation{\CardiffAddress}
\author{R. Simmons}       \affiliation{\AtiAddress}
\author{B.N. Murdin}    \affiliation{\AtiAddress}
\author{U. Zeitler}     \affiliation{\HFMLAddress}\affiliation{\IMM}
\author{P.D. Buckle}    \affiliation{\CardiffAddress}

\date{\today}
\begin{abstract}
We studied InSb quantum well devices using Landau level tunneling spectroscopy through a three-terminal differential conductance technique. This method is similar to filled state scanning tunneling microscopy but uses a stationary contact instead of a mobile tip to analyze the two-dimensional electron system. Applying magnetic fields up to 15~T, we identified clear peaks in the differential current-voltage profiles, indicative of Landau level formation. By examining deviations from the expected Landau fan diagram, we extract an absolute value for the exchange-induced energy shift. Through an empirical analysis, we derive a formula describing the exchange shift as a function of both magnetic field strength and electron filling. Our findings indicate that the emptying of the $\nu=2$ and $\nu=3$ Landau levels causes an exchange interaction energy shift in the $\nu=1$ level. Unlike prior studies that infer level energies relative to one another and report oscillatory g-factor behavior, our method references the energy of the Landau levels above the filled states of the contact under a bias voltage, revealing that only the ground state Landau level experiences a measurable exchange shift. 
\end{abstract}

\maketitle

\section{Introduction}

The exploration of quantum behavior in two-dimensional semiconductors subjected to strong magnetic fields has been a prominent research focus since the identification of the quantum Hall effect\cite{Klitzing1980}. Among these studies, the spectroscopic analysis of Landau levels has generated significant interest, particularly the exchange enhancement phenomenon\cite{Nicholas1988, Main2000, Ando1974}. In such systems, the electron energy is characterized by a combination of kinetic energy and exchange energy, the latter originating from the exchange interactions between electrons across different Landau levels.

The enhancement of the effective g-factor as a function of the external magnetic field has been widely reported and is commonly attributed to the exchange interaction between electrons \cite{Masutomi2015, Mu2016, Tang2011, Bovkun2015, Krishtopenko2012, Hayne1992, Krishtopenko2015, Nicholas1988, Nedniyom2009, Raymond1985, Krishtopenko2012_2, Savelev1996, Haug1988, Krishtopenko2011, Main2000}. This phenomenon was first investigated by Fang and Stiles in 1968 \cite{Fang1968}, who observed a variation in the g-factor with surface electron concentration in Si inversion layers under tilted magnetic fields.
This tilted field enhances Zeeman splitting while preserving Landau quantization, which depends only on the perpendicular field component.
The enhancement of the g factor is quantified by identifying the field orientation in which adjacent Landau levels align.
This method has since been widely adopted in the study of low-dimensional systems and for the understanding of many-body effects in quantum wells.

This enhancement in the exchange interaction is commonly linked to the polarization of electron spin of the spin-split Landau levels. Based on the findings of Fang and Stiles\cite{Fang1968}, Ando and Uemura\cite{Ando1974} predicted an oscillation in the g-factor originating from the electron spin polarization when the Landau levels intersect the Fermi energy. The nature of g-factor measurements is that they are relative measurements of energy splitting between two Landau levels, either using the tilted-field co-incidence\cite{Brosig2000,Fang1968,Yarlagadda1991,Tang2011,Haug1988,Giuliani1985,Lei2020,Gammag2013} or optical spectroscopic measurements\cite{Goldberg1989,Pinczuk1992}.
Reports of oscillating g-factors in III-V semiconductors\cite{Raymond1985,Sadofyev2002,Aleshkin2008} stem partly from flawed Shubnikov-de Haas (SdH) analysis methods\cite{Nizhankovskii1988,Nizhankovskii2011} that misattribute periodicity shifts to $g^*$ variations.

In this study, we report on three-terminal differential conductance measurements, which enables a direct measurement of Landau levels and the exchange enhancement effect. This method is similar to an earlier report of Landau-level spectroscopy of a GaAs surface two-dimensional system, using tunnnelling through an InAs/AlAs quantum dot\cite{Main2000}. However, this study has been carried out in a $\delta$-doped quantum well system where tunnelling is through a delta-doped layer rather than a single zero-dimensional ground state in a quantum dot. Another distinction lies in the study of an InSb/AlInSb quantum well (QW), where the InSb 2D system exhibits the largest bulk $g^*$ and the smallest effective mass ($m^*$) among the III-V binary compounds. These material properties leads to a rapid separation of the Landau levels into clearly identifiable Zeeman split spin states at comparatively low magnetic fields. Thereby allowing a clear and absolute measurement of the exchange interaction in this system.

\section{Experimental Methods}

\begin{figure}
    \centering
    \includegraphics[width=\linewidth]{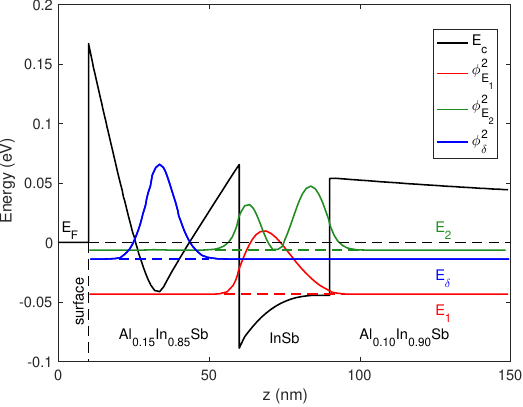}
    \caption{The zero bias band-structure and wavefunction solutions of the InSb/InAlSb QW, calculated using nextnano's Schrodinger-Poisson solver. The Fermi energy $E_\text{F}=0$.}
    \label{fig:structure}
\end{figure}

Experiments were performed on material grown by solid source MBE on semi-insulating, lattice-mismatched GaAs substrates.  In growth order, the epitaxy comprises an aluminium antimonide (AlSb) accommodation layer, a $3~\mu$m Al$_{0.1}$In$_{0.9}$Sb strain-relieving barrier layer (to allow for lattice mismatch relaxation), a $30$~nm InSb quantum well layer and a $50$~nm Al$_{0.15}$In$_{0.85}$Sb top barrier layer.  Tellurium (Te) $\delta$ doping is introduced only into the top barrier, $25$~nm above the InSb quantum well.  Deliberate doping of the lower barrier is avoided in order to prevent any impurity donor atoms from being carried forward on the growth plane which could significantly compromise the transport lifetime of carriers in the quantum well. The quantum wells have been modelled using the Schr\"odinger-Poisson software Nextnano\cite{Nextnano}, shown in \cref{fig:structure}, to determine subband states confined within the QW.  The $\Gamma$-point of the first subband $E_1$ is located 43~meV below the Fermi energy $E_\text{F}=0$, and indicates occupation of the second subband and states within the $\delta$-doped region. 
The energy of the conduction band at the surface is set to be 1/3 ($\approx0.17$~eV) of the Al$_{0.15}$In$_{0.85}$Sb bandgap ($\approx0.51$~eV) above $E_\text{F}$.
Note that this Schrod\"inger-Poisson model excludes the effects from the diffusion of Zn under the metal contact.

Devices were fabricated using standard photolithography techniques into six contact hall bars with a nominal distance of 200~$\mu$m between longitudinal contacts and 40~$\mu$m between transverse contacts. To form ohmic contacts, a Zn keying layer (around 10 nm) is initially deposited to improve adhesion to the surface followed by a thick Au layer (300 nm). These are evaporated while the substrate is heated to
$100 \si{\degree}$C to promote diffusion of the metal ions on the surface and minimise Schottky contact formation.
SdH measurements were performed to determine the low-temperature carrier concentration $n=3.14\times10^{15}$~m$^{-2}$, and to estimate the broadening of the Landau levels and the $|g^*|$, see supplementary materials.
In this investigation, a three-terminal differential conductance measurement was carried out on an InSb/InAlSb heterostructure, see \cref{fig:IV}a.
The three-terminal geometry isolates tunneling between the 2DEG and the contact by localizing the voltage drop at the common contact.
A further explanation of this geometry is provided in the Supplementary Materials. Using a force bias voltage with a 3~mV modulation, a software lock-in method was used to simultaneously record both the three-terminal differential conductance and I-V measurements of the InSb/InAlSb QW devices in a liquid helium bath at 4 K in fields up to 15~T.

\section{Results}
\begin{figure}
    \centering
    \includegraphics[width=\linewidth]{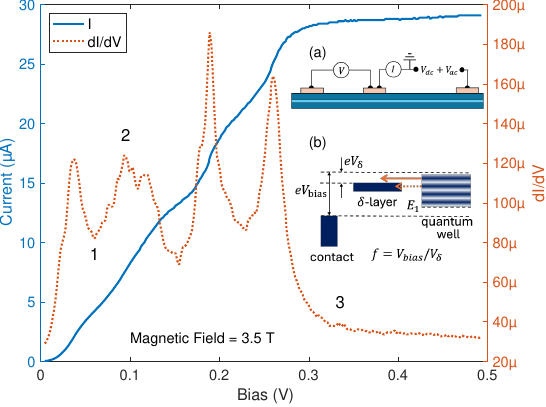}\\
    \caption{A typical three-terminal conductance measurement at 3.5~T, with peak-to-peak a.c. voltage $V_{ac}=~2$~mV at 33~Hz. The solid line shows the I-V measurement which aligns with the peaks seen in the dI/dV dotted line. (1) minima in dI/dV, corresponding filled upper states of $\delta$-layer aligned between Landau levels, (2) peak in dI/dV, corresponding to an increase in tunnelling as a Landau Level passes through the filled upper states of $\delta$-layer and can tunnel into empty states, (3) saturation, corresponding to the state where all Landau Levels can tunnel into empty states.  (a) Inset shows circuit diagram of three-terminal measurement. (b) Schematic of tunneling from Landau levels in quantum well through $\delta$-layer sub-band into metal contact under bias. The condition shown corresponds to the minima (1) in $dI/dV$ at which the upper filled states in the $\delta$-layer lies between the  $\nu=3$ and 4 Landau levels, such that tunnelling from the $\nu=3$ level is blocked.}
    \label{fig:IV}
\end{figure}

A typical I-V and differential conductance plots are shown in \cref{fig:IV} for forward bias in a magnetic field of 3.5~T. The I-V characteristics exhibit a stepwise increase in current with increasing bias voltage, which saturates at approximately 0.3~V with a current of 28~$\mu$A. These steps are manifested as peaks in the differential conductance. 
This is analogous to filled-state scanning tunneling microscopy (STM), where filled states tunnel into empty states in the tip\cite{Tersoff1985}. A study using this STM technique has reported Landaus level spectroscopy of a Cs-induced inversion layer on the surface of InSb\cite{Becker2011}, in which a small oscillation was observed in the exchange interaction for high filling factors. In this study, the filled states in the QW tunnel into empty states above the Fermi energy ($E_\text{F}$) in the metal contact. In the example shown in \cref{fig:IV} at 3.5~Tesla, there are four filled Landau levels below $E_\text{F}$. As the bias voltage is increased the upper filled Landau level is raised above the Fermi level of the metal contact (region 1), allowing electrons to tunnel into the empty states giving rise to a step increase in the current and the first peak in the differential conductance. Increasing the bias further brings more Landau levels above the Fermi level of the metal contact, where peaks in the differential conductance correspond to the condition where the Landau level passing through the Fermi level of the metal contact (region 2), allowing the electrons to tunnel into empty states. Eventually, all of the Landau levels, i.e. all the filled states in the QW, are raised above the Fermi level of the metal contact such that the current saturates (region 3).

\begin{figure}
    \centering
    \includegraphics[width=\linewidth]{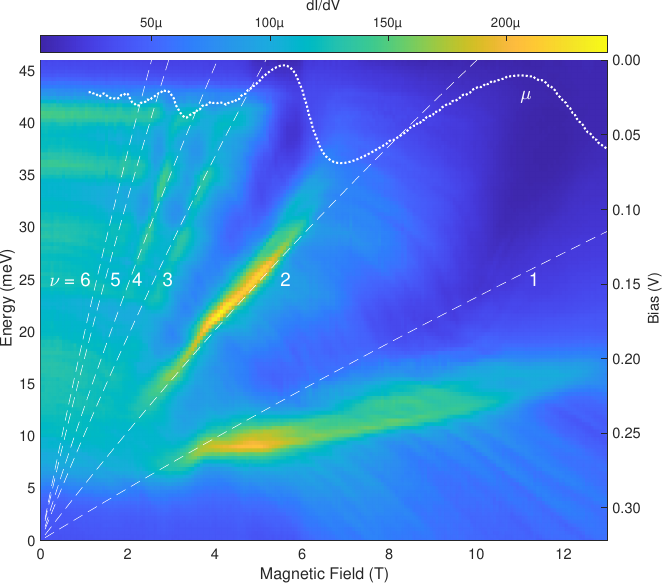}
    \caption{Differential conductance as a function of bias and magnetic field. The calculated energies of Landau level fan plot overlaid on the differential conductance data, as determined through fitting to the extracted peaks in the differential conductance for the first four Landau levels $\nu=1 (l=0,\lambda=+1),\nu=2(l=1,\lambda=-1),\nu=3(l=1,\lambda=+1)$, $\nu=4(l=2,\lambda=-1)$, $\nu=5(l=2,\lambda=+1)$, and $\nu=6(l=3,\lambda=-1)$. The bias is converted to Landau level energy using fitting parameters $E_1$ and $f$. The calculated electrochemical potential $\mu$ is also shown, see supplementary materials.}
    \label{fig:heatmap}
\end{figure}

A differential conductance intensity plot as a function of applied magnetic field and bias voltage is shown in \cref{fig:heatmap} which exhibits a Landau level fan with the first four spin-split levels labeled with filling factor $\nu$. The additional texture in the Landau fan plot is due to density of states of the contact, since the tunneling current is a product of the density of states of the quantum well and the contact. At higher magnetic fields above 5~T, oscillations in differential conductance which have the opposite field dependence to the Landau levels are observed. These features can be explained by the formation of Landau levels in the delta layer.
Above 4~T, the $\nu=1$ Landau level shows a noticeable downward shift due to exchange, labeled $E_\text{ex}$, which we quantify by fitting the Landau fan.
To model the Landau fan plot, which includes the spin-orbit interaction, we adopt the formalism of Bychkov and Rashba\cite{Bychkov1984,Hernangomez-Perez2013}, such that

\begin{equation}
    E=E_{l,\lambda}=\hbar\omega_c\left[l-\frac{\lambda}{2}\sqrt{\left(1-Z\right)^2+lS^2}\right],
    \label{Eq:BanrR}
\end{equation}

where $l=0,1,2,\dots$ is a positive integer, $\lambda=\pm1$ ($\lambda=+1$ only for $l=0$) and $\omega_\text{c}=eB/m^{*}(E)$ is the cyclotron frequency, $B$ is the perpendicular applied magnetic field and $m^{*}(E)$ is the effective mass which includes the effects of non-parabolocity as the band energy dispersion of InSb is highly non-parabolic due to its small bandgap. The Rashba spin-orbit and Zeeman terms are given by

\begin{equation}
    S=\frac{\alpha\sqrt{2}}{\hbar\omega_c l_B},\quad Z=\frac{g^*(E)m^{*}(E)}{2m_0},
    \label{Eq:SandZ}
\end{equation}

where $\alpha$ is the Rashba parameter, $l_\text{B}=\sqrt{\hbar/eB}$ is the magnetic length, $m_0$ is the free electron mass, and $g^*(E)$ is the g-factor which also includes the effects of non-parabolicity. 
We obtain the theoretical estimates of both the effective mass $m^*(\Gamma)=0.013m_0$ and g-factor $g^*(\Gamma)=-47$ at the $\Gamma$-point from $\mathbf{k}\cdot\mathbf{p}$ theory\cite{Ihn2009,Hermann1977}, see supplementary materials.
To incorporate the non-parabolicity into the Landau fan plot calculation, an approximation in the limit where $m^*(\Gamma)\ll1$ and $g^*(\Gamma)\gg2$ is used, such that
\begin{equation}
    m^*(E)\approx m^{*}(\Gamma)(1+\Lambda E),
    \label{Eq:Lambda}
\end{equation}
\begin{equation}
    g^*(E)\approx \frac{g^*(\Gamma)}{1+\Lambda E}.
\end{equation}
where $\Lambda\approx1/(E_0+E_\text{c})=4.0$~eV$^{-1}$.

We see that the Zeeman term $Z$ is unaffected by non-parabolicity due to the product $m^*g^*$ being independent of $\Lambda$.
The Rashba term $S$ is also is unaffected by non-parabolicity as both $\omega_\text{c}$ and the Rashba parameter $\alpha$ scale inversely with $m^*$\cite{Andrada1994}.
The Landau energies including non-parabolicity is therefore

\begin{equation}\label{Eq:LL_energy}
    \Lambda E^2+E-\frac{\hbar eB}{m^*(\Gamma)}\left[l-\frac{\lambda}{2}\sqrt{\left(1-Z\right)^2+lS^2}\right]=0.
\end{equation}

The Landau level energies can be extracted by taking the positive solutions to this quadratic equation.
To fit a Landau fan to the data, there are now four material-specific fitting parameters which include $m^*(\Gamma)$, $|g^*(\Gamma)|$, $\Lambda$ and $\alpha$.
To reduce the number of free fitting parameters the values of $\Lambda=4.0$~eV$^{-1}$ and $\alpha=0.1$~eV\AA\cite{Li2012} were fixed.
The latter has minimal effect on the Landau fan plot at the fields used in this study.

Robust fitting was found with $|g^*(\Gamma)|=50$, $m^*(\Gamma)=0.0145m_0$, $E_1=-47$~meV using a leverage factor $f=7.0$~V/eV which is defined as the bias voltage required to produce a relative energy shift of $E_1$ with respect to the upper filled state in the $\delta$-layer, see \cref{fig:IV} inset (b). See Supplementary Materials for more details of the fitting process. 
The value of $m^*(\Gamma)$ is higher than predicted from $\mathbf{k\cdot p}$ theory, although it is within the experimentally measured range. The agreement of the fitting parameters with those predicted from theory, see \cref{tab:compare} for summary, provides confidence in the value of the obtained leverage factor, which ultimately determines the magnitude of the observed shift in energy of the first Landau level above 4~T.
We discount the possibility of a field dependence on the leverage factor as the fit to the Landau fan plot for $\nu=2$ maintains good accuracy well beyond 4~T, past the stage at which the shift in the $\nu=1$ level becomes evident.
\begin{table}
    \centering
    \begin{tabular}{c|c|c}
        Parameter & Theoretical/modeled & Experimental fit\\ \hline\hline
        $m^*(\Gamma)$ & 0.0131$m_0$ & 0.0145$m_0$ \\
        $|g^*(\Gamma)|$ & 47 & 50\\
        $E_1$ & -43~meV  & -46~meV\\
        leverage factor $f$ & - & 7.0 V/eV \\
    \end{tabular}
    \caption{Comparison of parameters obtained from theoretical values or through modeling with those obtained from the experimental fits to the Landau fan accounting for non-parabolicity $\Lambda=4.0$~eV$^{-1}$.}
    \label{tab:compare}
\end{table}

\begin{figure}
    \centering
    \includegraphics[width=0.95\linewidth]{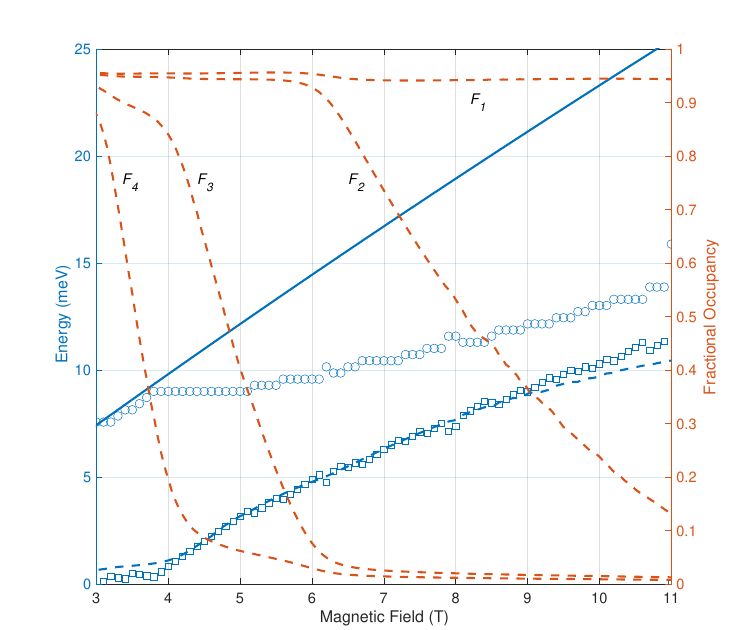}
    \caption{The energy and exchange energy shift for the ground state $\nu_1$ Landau level. Solid line: calculated Landau level energy for $\nu=1$, circles: measured $\nu=1$, squares measured $E_\text{ex}$, dotted line: fit to $E_\text{ex}$ using \cref{Eqn:Eex_fit}. Also shown are the calculated fractional electron occupation of the first four Landau Levels plotted as a function of the magnetic field at 4~K. }
    \label{fig:E_ex}
\end{figure}

The measured shift in energy $E_\text{ex}$ of the $\nu=1$ level is shown in \cref{fig:E_ex} as a function of the magnetic field. This was calculated by taking the difference between the sampled peak values in the measured data from the expected Landau level energy using the best-fit parameters in \cref{tab:compare}. In a magnetic field of 11~T the observed shift is approximately 11~meV, at a field where the majority of electrons occupy the ground state Landau level. The relationship between this exchange energy shift and the electron occupation of Landau levels will be discussed in the next section.


\section{Discussion}\label{section:discussion}

Previously, the shift in energy due to the exchange enhancement is assumed to be proportional to the polarization of the ground-state Landau level\cite{Nicholas1988}.
We adopt a more nuanced approach taking into account the exchange interaction between electrons in the ground state Landau level ($\nu=1$) and Landau levels with higher filling factors ($\nu>1$).
In order to investigate this, the following equation was used to define the strength of $E_\text{ex}$

\begin{equation}
    E_\text{ex}=\sum_{\nu=2}E^\nu_{\text{ex}}(1- F_{\nu}),
    \label{Eqn:Eex_fit}
\end{equation}

where $E^\nu_\text{ex}$ are coefficients representing the product of the Coulomb interaction and the exchange integral between the ground state Landau level and the Landau level with filling factor $\nu$, and $F_\nu$ is the fractional electron occupation of the higher Landau level.

The fractional occupations were determined by numerically solving the broadened Landau level density of states at finite temperature (details in Supplementary Material).
Plotting these electron occupations shows that the $\nu=4$ level starts to deplete at approximately 3~T before any detectable shift in the $\nu=1$ level occurs, see \cref{fig:E_ex}. Notably, this shift aligns with the onset of the depletion for the $\nu=3$ level at around 4~T.


The expression in \cref{Eqn:Eex_fit} is fitted using a linear regression to obtain the coefficients of $E_{\text{ex}}^2=6.6$~meV and $E_{\text{ex}}^3=4.6$~meV, with negligible values for $E_{\text{ex}}^4$ and above, with the fit obtained shown in \cref{fig:E_ex}.
This empirical, phenomenological method of investigating the behavior of electrons in a 2DEG quantum well provides evidence that the $\nu=2$ and $\nu=3$ levels have similar interaction strengths with the $\nu=1$ level.
The slight deviation between the measured exchange shift and the fit of $E_\text{ex}$ could be attributed to a further reduction in the coulomb interaction between electrons in $\nu=1$ level as they become more localized due to the reduction of the magnetic length which is proportional to $1/\sqrt{B}$.

\section{Conclusions}
We have presented an absolute measurement of the Landau fan in InSb quantum wells using a three-terminal differential conductance technique. This method enables direct investigation of the electronic behavior in the presence of large external magnetic fields.
Unlike conventional approaches that determine Landau level energies relative to other field-dependent levels, our technique leverages a relative linear shift in the states in the Landau levels with the upper filled states in the $\delta$-layer barrier. This allows for the measurement of the level energies independently of other levels.
Our results reveal a distinct deviation of the ground state $\nu=1$ Landau level from the expected fan diagram, a deviation attributed to the exchange interaction.
Interestingly, this spectroscopic technique does not indicate similar deviations for higher Landau levels ($\nu > 1$), contradicting previous reports of oscillatory behavior in the effective g-factor.
A potential explanation for this observation is rooted in the spatial distribution and spin configuration of the Landau level wavefunctions. 
At high fields, the reduced cyclotron radius promotes spin polarization and spatial separation, suppressing Coulomb repulsion in the $\nu = 1$ level.
This limits the self-exchange energy in the ground state. As the magnetic field decreases and higher Landau levels become occupied, strong exchange interactions arise between the $\nu = 1$ level and partially filled levels of the same spin. In particular, the $\nu = 3$ level, which shares the same spin orientation as $\nu = 1$, couples strongly via exchange. As $\nu = 3$ depopulates, this strong coupling enables electrons in both levels to reconfigure, thereby reducing their mutual Coulomb interaction and leading to a measurable shift in the energy of the ground state. The $\nu = 2$ level, although of opposite spin, also contributes to the exchange shift via Coulomb interaction, and its depletion similarly reduces the interaction strength. In contrast, levels such as $\nu = 4$ are either spin-paired or more weakly interacting due to reduced overlap or screening, and thus have negligible influence. These effects explain why only the $\nu = 2$ and $\nu = 3$ levels significantly affect the exchange shift of the $\nu = 1$ Landau level.
\\
In summary, these findings highlight that the exchange interaction selectively affects the ground-state spin-split Landau level and does not manifest as a global modification of the g-factor.
Thus, describing the exchange shift as a renormalized g-factor is misleading; the interaction is independent of spin magnetic moment and better treated as a distinct exchange effect.


\section*{Acknowledgments}
This work was supported by HFML-RU/NWO-I, member of the European Magnetic Field Laboratory (EMFL). The authors warmly remember the late Dr. Phil Buckle (d. 2021) and express their heartfelt thanks to his wife, Dr. Louise Buckle, for her generous permission to see this paper published posthumously.

\setcounter{figure}{0}
\renewcommand{\thefigure}{S\arabic{figure}}
\renewcommand{\theequation}{S\arabic{equation}}
\setcounter{equation}{0}
\section*{Supplementary Materials}
\subsection*{Shubnikov-de Haas Measurements}
\begin{figure}
    \centering
    \includegraphics[width=0.8\linewidth]{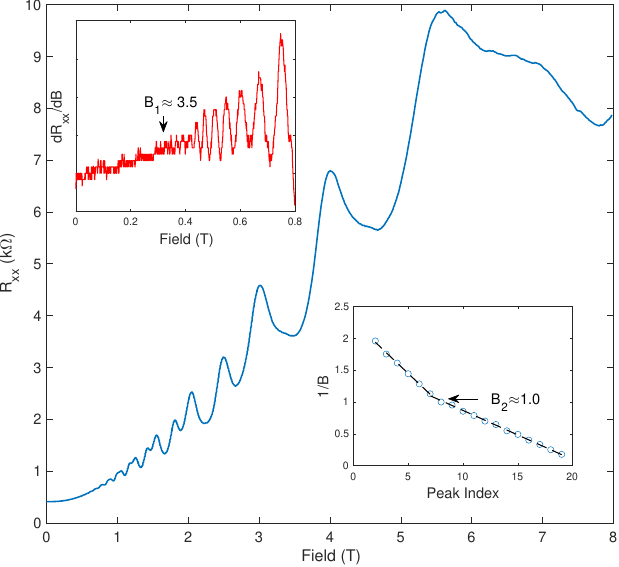}
    \caption{Shubnikov-de Haas measurement of the $\delta$-doped InSb/AlInSb. Upper left inset: $dR_{xx}/dB$ showing onset of oscillations at $B_1=0.35$~T. Lower right inset: Peak position ($1/B$) versus peak index. The change in gradient at $B_2=1$~T indicates the point at which the Landau levels are spin resolved. }
    \label{fig:peak_index}
\end{figure}

The charge carrier concentration of the $\delta$-doped InSb/AlInSb was determined through the Shubnikov-de Haas (SdH) transport measurement at $T=4$~K shown in \cref{fig:peak_index}. Key fields are highlighted: $B_1$ represents the threshold field for observing SdH oscillations which is determined by the onset of oscillations in the derivative $dR_{xx}/dB$, as seen in upper left inset in \cref{fig:peak_index}; $B_2$ marks the field where spin-splitting becomes discernible, as seen in the lower right inset in in \cref{fig:peak_index} which presents the SdH peak index plotted against $1/B$. The lower right inset enables the extraction of the sheet carrier density from the gradient, applying the formula
\begin{equation}
    \Delta(1/B)=\frac{2e}{hn_s},
\end{equation} 
for the region $B_1<B<B_2$ and
\begin{equation}
    \Delta(1/B)=\frac{e}{hn_s},
\end{equation}
where $B>B_2$ in the region where the spin-splitting is resolved. Using the gradient for $B>1$~T, a carrier concentration $n_s=3.14\times10^{15}$~m$^{-2}$ is obtained.\\

The Landau level broadening can be estimated from the condition $\Gamma=\hbar\omega_c$ at the onset of the observation of Shubnikov-de Hass oscillations. This is expressed using the formula 
\begin{equation}
    \Gamma=\frac{\hbar e B_1}{m^*},
\end{equation}
where $B_1=0.35$~T and $m^*\approx0.0145m_e$, giving $\Gamma\approx2.8$~meV. Similarly, at the critical field $B_2$, the spin-splitting is comparable to the Landau level broadening
\begin{equation}
    g*\mu_\text{B}B_2\approx\Gamma\approx\frac{\hbar e B_1}{m^*},
\end{equation}
so that
\begin{equation}
    g^*\approx\frac{\hbar e}{\mu_\text{B}m^*}\frac{B_1}{B_2}.
\end{equation}
With $B_2=1$~T, we obtain $g^*\approx48$. These estimates obtained from the SdH data are consistent with the parameters obtained through the fitting of the experimentally obtained Landau fan plot.

\subsection*{Three-terminal Measurements}

\begin{figure}
    \centering
    \includegraphics[width=0.7\linewidth]{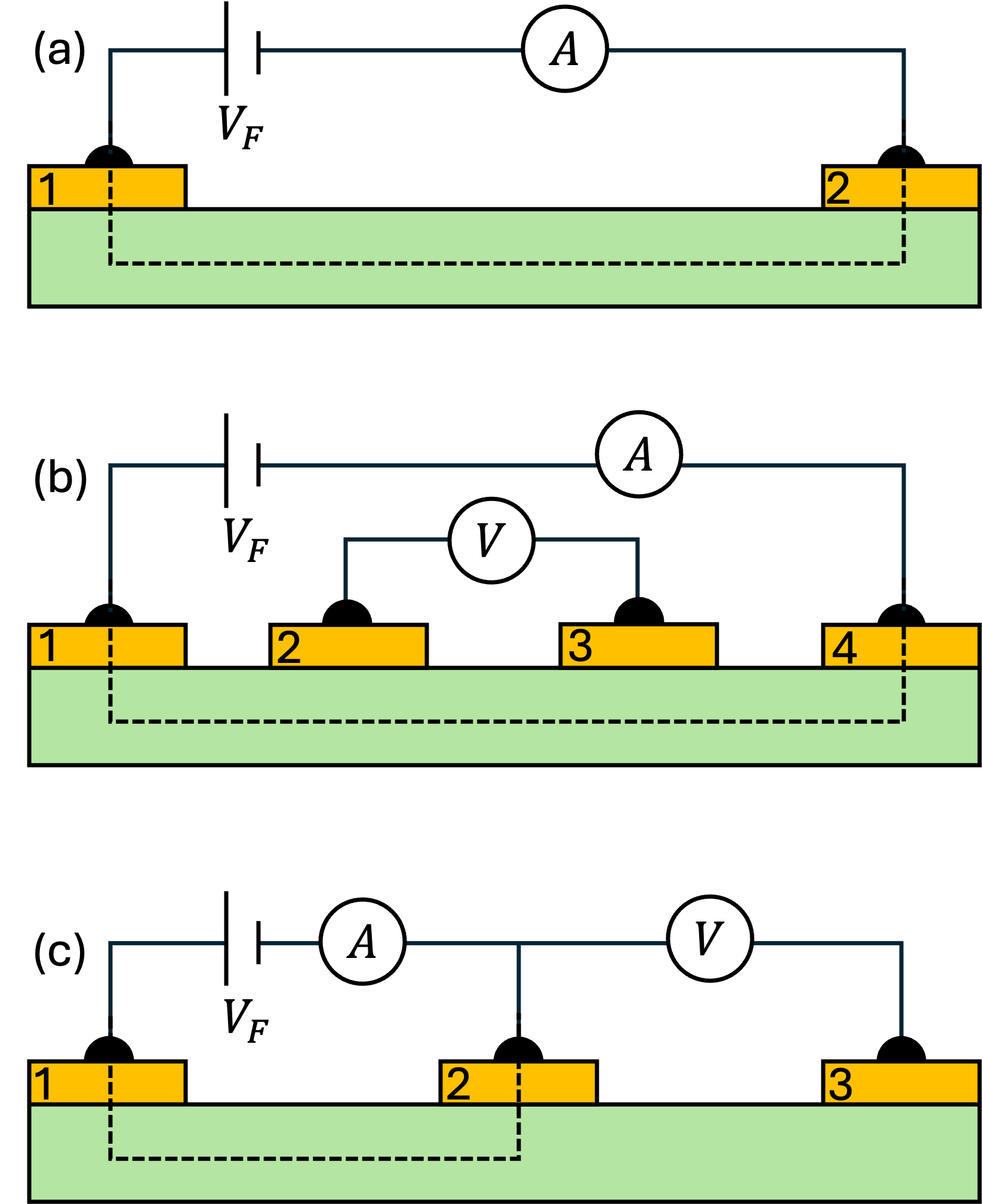}
    \caption{Geometry of (a) two-terminal, (b) four-terminal and (c) three-terminal measurements. In these geometries, a force voltage $V_\text{F}$ is applied and current and voltage potential differences are measured. The dotted line indicates the current path through the circuit.}
    \label{fig:3_term}
\end{figure}

The three-terminal electrical measurement configuration may be less well-known compared to the conventional two- and four-terminal setups. Since this three-terminal setup is crucial for the measurements in our study, we describe it here in relation to the two- and four-terminal geometries. The two-terminal configuration is depicted in \cref{fig:3_term}(a), where a driving voltage $V_\text{F}$ is applied across a pair of contacts, and the resulting electrical current $I$ is observed. The total resistance
\begin{equation}
    R=\frac{V_\text{F}}{I}=R_\text{c}+R_\text{s}+R_\text{c}
\end{equation}
measured includes $R_\text{c}$, the resistance of both contacts, and $R_\text{s}$, the resistance of the substrate between contacts 1 and 2. 
The four-terminal geometry is often employed to eliminate contact resistance effects from measurements, as demonstrated with the Shubnikov-de Haas measurement previously. In \cref{fig:3_term}(b), the driving voltage $V_\text{F}$ is applied between the outer terminals, 1 and 4, while the voltage is recorded across the inner terminals, 2 and 3. Importantly, there is no current passing through the inner contacts, resulting in no potential drop across them. Consequently, the recorded voltage is exclusively due to the potential drop across the substrate between contacts 2 and 3. So that the resistance measured is
\begin{equation}
    R=\frac{V}{I}=R_\text{s}.   
\end{equation}
\\
In the three-terminal configuration, a drive voltage is applied between outer contact 1 and the central contact 2. The voltage is subsequently measured between the central contact and the outer contact 3. This setup ensures that no current passes through contact 3 or the substrate between contacts 2 and 3, making the voltage measurement solely sensitive to the potential difference across contact 2. So that
\begin{equation}
   R=\frac{V}{I}=R_\text{c}, 
\end{equation}

where $R_{c}$ is the resistance of contact 2. This configuration enables the investigation of the V characteristic of the resistance of a single contact, attributed in this study to the vertical tunneling of the electrons from the quantum well, through the $\delta$ layer, to the metal contact.

\subsection*{Theoretical $\mathbf{k}\cdot\mathbf{p}$ calculations}

As a guide to the fitting procedure, we can obtain theoretical estimates of both the effective mass $m^*(\Gamma)$ and g-factor $g^*(\Gamma)$ at the $\Gamma$-point, which can be extracted from $\mathbf{k}\cdot\mathbf{p}$ theory\cite{Ihn2009,Hermann1977} given by


\begin{equation}
    \frac{m_\text{0}}{m^*(\Gamma)}=1+\frac{2m_\text{e}P^2}{3\hbar^2}\left( \frac{2}{E_\text{0}+E_\text{c}}+\frac{1}{E_\text{0}+\Delta_0+E_\text{c}} \right)-\frac{2m_\text{e}P'^2}{3\hbar^2}\left(\frac{2}{E'_0-E_0+\Delta'_0}+\frac{1}{E'_0-E_0} \right)
    \label{Eq:m*}
\end{equation}
     
\begin{equation}
    g^*(\Gamma)=2-\frac{4m_\text{e}P^2}{3\hbar^2}\left( \frac{1}{E_0+E_\text{c}}-\frac{1}{E_0+E_\text{c}+\Delta_0}\right) +\frac{4m_\text{e}P'^2}{3\hbar^2}\left( \frac{1}{E'_0-E_0}-\frac{1}{E'_0-E_0+\Delta'_0}\right),
    \label{Eq:g*}
\end{equation}

which accounts for the quantum well confinement energy\cite{Lei2020} which has been calculated from the Nextnano simulation using $E_\text{c}=\int{\phi^2_{E_1}(z)E_\text{c}(z)dz}=14$~meV.
We have adopted band structure parameters used by ref\cite{Winkler2003} of $E_0=0.237$~eV, $E'_0=3.160$~eV, $\Delta_0=0.810$~eV, $\Delta'_0=0.330$~eV, $P=9.641$~eV\AA, and $P'=6.324i$~eV\AA. This gives the theoretical values of $m^*(\Gamma)=0.013m_0$ and $g^*(\Gamma)=-47$.

\subsection*{Fitting to Landau Fan Plot}

\begin{figure}[htbp]
    \centering
    \includegraphics[width=0.5\linewidth]{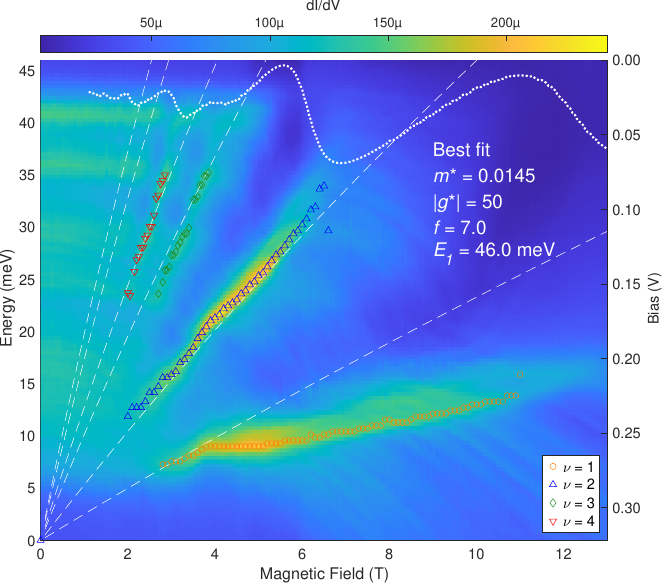}\\
    \includegraphics[width=0.5\linewidth]{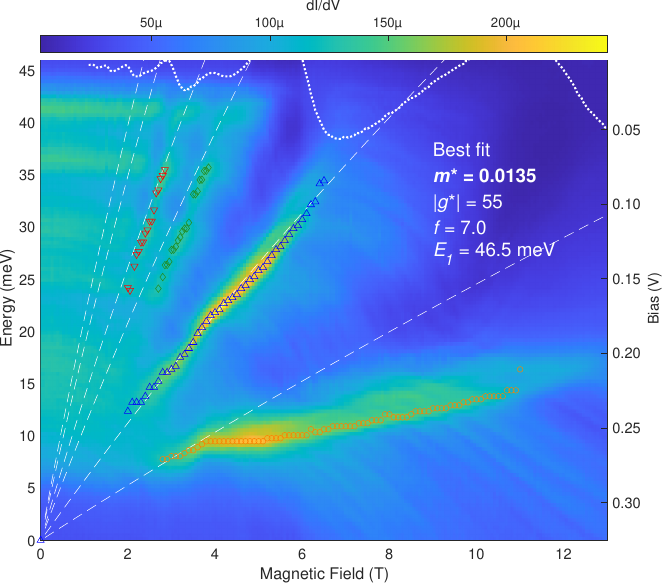}\\
    \includegraphics[width=0.5\linewidth]{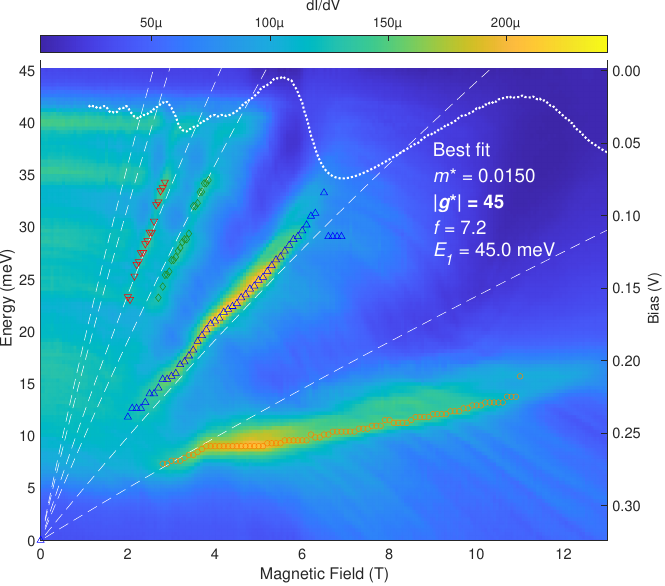}
    \caption{Comparison of fitting results. (Upper panel) best fit with parameters used in study, (middle panel) best fit with $m^*=0.0135$, and (lower panel) best fit with $|g^*|=0.45$.}
    \label{fig:fitting_combined}
\end{figure}

The fitting process for the Landau fan plot involved identifying the peak positions of the first four Landau levels. For $\nu=1$ and $\nu=2$, this was automated, while $\nu=3$ and $\nu=4$ were manually determined from the raw I-V curves. Optimal fitting was achieved by manually tweaking the parameters, guided by the predicted values from the $\mathbf{k\cdot p}$ model. Note that data points for $\nu=1$ above 4~T were excluded from the fitting process, as the shift in $\nu=1$ is observed beyond this threshold. The fit for $\nu=5$ additionally serves as a visual reference, confirming that the energy gap between $\nu=4$ aligns with the fit.\\
To demonstrate the robustness of the fit, the best fits with slightly adjusted values of $m^*$ and $|g^*|$ are shown in \cref{fig:fitting_combined}. For the case where $m^*=0.0135$ is fixed, the best-fit is shown in the middle panel in 
\cref{fig:fitting_combined}, where the model fits well to $\nu=1$ and $\nu=2$ but is worse for $\nu=3$ and $\nu=5$. It is also noted that the calculated electrochemical potential does not align with point at which the differential conductance drops in intensity. The lower panel in \cref{fig:fitting_combined} shows the best fist in the case where $|g^*|=45$ is fixed. Here we obtain a good fit to all but the $\nu=2$ Landau level. Further deviations from the optimized parameters lead to a worsening in the resulting fit. The potential for the $\nu=2$ Landau level to undergo an exchange shift that leads to a reduction in energy beyond 4~T was thoroughly investigated. However, aligning this model with the data proved challenging, and a satisfactory fit was not achieved.

\section*{Calculation of Fractional Occupation of Landau Levels}

To determine the fractional electron occupation of the Landau levels, the broadened Landau level density of states at finite temperature is used to numerically calculate the electrochemical potential. This in turn allows the populations of each of the Landau levels to be calculated. The first step is to calculate the total density of states in an applied field due to the contribution of all Landau levels, which is described by

\begin{equation}
	g\left(E,B\right)=\frac{eB}{h}\sum_{l,\lambda}{\frac{\Delta E}{2\pi\left(E-E_{l,\lambda}\right)^2+	\left(\frac{\Delta E}{2}\right)^2}},
	\label{Eqn:DOS}
\end{equation}

where $\Delta E\approx2.8$~meV is the Lorentzian broadening of the Landau levels, which has been estimated from the observed onset of the Shubnikov de Haas oscillations. This broadening is in good agreement with Lorentzian peak widths of Landau levels in I-V curves, which are $\sim35$~mV corresponding to $\sim5$~meV using the leverage factor 7.0~V/eV.  It is now possible to determine the electrochemical potential in a particular field by numerically satisfying the condition.

\begin{equation}
	\int_0^\mu f\left(E,\mu,T\right)g\left(E,B\right)\text{d}E=n,
    \label{Eqn:Chemical_Potential}
\end{equation}

where $\mu$ is the electrochemical potential, $f(E,\mu,T)$ is the Fermi function and $n$ is the sheet electron density which has been determined from SdH measurements. The calculated $\mu$ as a function of the magnetic field is seen to oscillate as Landau levels are emptied, as shown in \cref{fig:fitting_combined}a . Following this numerical solution to obtain $\mu$, \cref{Eqn:Chemical_Potential} can then be applied to individual Landau levels to obtain the populations $n_\nu$ of each Landau level at a given $B$.  The electron occupation for each filling factor can be extracted by taking the population, $n_\nu$ and dividing this by the Landau Level density of states, $F_\nu = n_\nu/(eB/h)$. This describes the relationship for the fraction of occupied states in each Landau Level as a function of the magnetic field.


\begin{thebibliography}{10}

\bibitem{Klitzing1980}
K.~v. Klitzing, G.~Dorda, and M.~Pepper.
\newblock New method for high-accuracy determination of the fine-structure constant based on quantized hall resistance.
\newblock {\em Phys. Rev. Lett.}, 45:494--497, Aug 1980.

\bibitem{Nicholas1988}
R~J Nicholas, R~J Haug, K~V Klitzing, and G.~Weimann.
\newblock {Exchange enhancement of the spin splitting in a GaAs-GaxAl1-xAs heterojunction}.
\newblock {\em Physical Review B}, 37(3):1294--1302, 1988.

\bibitem{Main2000}
P.~C. Main, A.~S.~G. Thornton, R.~J.~A. Hill, S.~T. Stoddart, T.~Ihn, L.~Eaves, K.~A. Benedict, and M.~Henini.
\newblock Landau-level spectroscopy of a two-dimensional electron system by tunneling through a quantum dot.
\newblock {\em Phys. Rev. Lett.}, 84:729--732, Jan 2000.

\bibitem{Ando1974}
Tsuneya Ando and Yasutada Uemura.
\newblock {Oscillation of Effective g Factor in MOS Inversion Layers under Strong Magnetic Fields}.
\newblock {\em Japanese Journal of Applied Physics}, 13(S2):329, jan 1974.

\bibitem{Masutomi2015}
Ryuichi Masutomi and Tohru Okamoto.
\newblock {Adsorbate-induced quantum Hall system probed by scanning tunneling spectroscopy combined with transport measurements}.
\newblock {\em Applied Physics Letters}, 106(25), jun 2015.

\bibitem{Mu2016}
Xiaoyang Mu, Gerard Sullivan, and Rui-Rui Du.
\newblock Effective g-factors of carriers in inverted inas/gasb bilayers.
\newblock {\em Applied Physics Letters}, 108(1):012101, 2016.

\bibitem{Tang2011}
Ning Tang, Kui Han, Fang-Chao Lu, Jun-Xi Duan, Fu-Jun Xu, and Bo~Shen.
\newblock {Exchange Enhancement of Spin-Splitting in AlxGa1-xN/GaN Heterostructures in Tilted Magnetic Fields}.
\newblock {\em Chinese Physics Letters}, 28(3), mar 2011.

\bibitem{Bovkun2015}
L.~S. Bovkun, S.~S. Krishtopenko, M.~S. Zholudev, A.~V. Ikonnikov, K.~E. Spirin, S.~A. Dvoretsky, N.~N. Mikhailov, F.~Teppe, W.~Knap, and V.~I. Gavrilenko.
\newblock {Exchange enhancement of the electron g-factor in a two-dimensional semimetal in HgTe quantum wells}.
\newblock {\em Semiconductors}, 49(12), dec 2015.

\bibitem{Krishtopenko2012}
S.S. Krishtopenko, V.I. Gavrilenko, and M.~Goiran.
\newblock {Exchange Enhancement of g-Factor in Narrow-Gap InAs/AlSb Quantum Well Heterostructures}.
\newblock {\em Solid State Phenomena}, 190, jun 2012.

\bibitem{Hayne1992}
M~Hayne, A~Usher, J.~J. Harris, and C~T Foxon.
\newblock {Exchange enhancement of the Landau-level separation for two-dimensional electrons in GaAs/Ga1-xAlxAs heterojunctions}.
\newblock {\em Physical Review B}, 46(15):9515--9519, 1992.

\bibitem{Krishtopenko2015}
S.~S. Krishtopenko, K.~V. Maremyanin, K.~P. Kalinin, K.~E. Spirin, V.~I. Gavrilenko, N.~V. Baidus, and B.~N. Zvonkov.
\newblock {Exchange enhancement of the electron g factor in strained InGaAs/InP heterostructures}.
\newblock {\em Semiconductors}, 49(2), feb 2015.

\bibitem{Nedniyom2009}
B.~Nedniyom, R.~J. Nicholas, M.~T. Emeny, L.~Buckle, A.~M. Gilbertson, P.~D. Buckle, and T.~Ashley.
\newblock {Giant enhanced g-factors in an InSb two-dimensional gas}.
\newblock {\em Physical Review B}, 80(12), sep 2009.

\bibitem{Raymond1985}
A.~Raymond, J.L. Robert, C.~Bousquet, W.~Zawadzki, F.~Alexandre, and I.M. Masson.
\newblock {Gigantic exchange enhancement of spin g-factor for two-dimensional electron gas in GaAs}.
\newblock {\em Solid State Communications}, 55(4), jul 1985.

\bibitem{Krishtopenko2012_2}
S.~S. Krishtopenko, K.~P. Kalinin, V.~I. Gavrilenko, Yu.~G. Sadofyev, and M.~Goiran.
\newblock {Rashba spin splitting and exchange enhancement of the g factor in InAs/AlSb heterostructures with a two-dimensional electron gas}.
\newblock {\em Semiconductors}, 46(9), sep 2012.

\bibitem{Savelev1996}
I~G Savel'ev, A~M Kreshchuk, S~V Novikov, A~Y Shik, G~Remenyi, Gy~Kov{\'{a}}cs, B~P{\^{o}}d{\"{o}}r, and G~Gombos.
\newblock {Spin splitting of the Landau levels and exchange interaction of a non-ideal two-dimensional electron gas in heterostructures}.
\newblock {\em Journal of Physics: Condensed Matter}, 8(46), nov 1996.

\bibitem{Haug1988}
R.J. Haug, K.~von Klitzing, R.J. Nicholas, J.C. Maan, and G.~Weimann.
\newblock {The influence of a tilted magnetic field on the fractional quantum hall effect and the exchange enhancement of the spin splitting}.
\newblock {\em Surface Science}, 196(1-3), jan 1988.

\bibitem{Krishtopenko2011}
S~S Krishtopenko, V~I Gavrilenko, and M~Goiran.
\newblock {Theory of g-factor enhancement in narrow-gap quantum well heterostructures}.
\newblock {\em Journal of Physics: Condensed Matter}, 23(38), sep 2011.

\bibitem{Fang1968}
F.~F. Fang and P.~J. Stiles.
\newblock Effects of a tilted magnetic field on a two-dimensional electron gas.
\newblock {\em Phys. Rev.}, 174:823--828, Oct 1968.

\bibitem{Brosig2000}
S.~Brosig, K.~Ensslin, and A.~Jansen.
\newblock Inas-alsb quantum wells in tilted magnetic fields.
\newblock {\em Physical Review B - Condensed Matter and Materials Physics}, 61:13045--13049, 2000.

\bibitem{Yarlagadda1991}
Sudhakar Yarlagadda.
\newblock Magnetization instabilities at tilted magnetic fields in the quantum hall regime.
\newblock {\em Physical Review B}, 44:13101, 1991.

\bibitem{Giuliani1985}
G.F. Giuliani and J.J. Quinn.
\newblock Spin-polarization instability in a tilted magnetic field of a two-dimensional electron gas with filled landau levels.
\newblock {\em Physical Review B - Condensed Matter and Materials Physics}, 31:6228--6232, 1985.

\bibitem{Lei2020}
Zijin Lei, Christian~A. Lehner, Km~Rubi, Erik Cheah, Matija Karalic, Christopher Mittag, Luca Alt, Jan Scharnetzky, Peter Märki, Uli Zeitler, Werner Wegscheider, Thomas Ihn, and Klaus Ensslin.
\newblock Electronic g factor and magnetotransport in insb quantum wells.
\newblock {\em Physical Review Research}, 2:33213, 2020.

\bibitem{Gammag2013}
Rayda Gammag and Cristine Villagonzalo.
\newblock Two-dimensional electron gas tilt-induced landau level crossings.
\newblock {\em Solid State Communications}, 156:16--20, 2013.

\bibitem{Goldberg1989}
B.~B. Goldberg, D.~Heiman, and A.~Pinczuk.
\newblock Exchange enhancement of a spin-polarized 2d electron gas determined by optical-absorption spectroscopy.
\newblock {\em Phys. Rev. Lett.}, 63:1102--1105, Sep 1989.

\bibitem{Pinczuk1992}
A~Pinczuk, B~S Dennis, '~D Heiman, C~Kallin, L~Brey, C~Tejedor, S~Schmitt-Rink, L~N Pfeiffer, and K~W West.
\newblock Spectroscopic measurement of large exchange enhancement of a spin-polarized 2d electron gas.
\newblock 68:24, 1992.

\bibitem{Sadofyev2002}
Yu~G Sadofyev, A~Ramamoorthy, J~P Bird, )~S~R Johnson, and Y.-H Zhang.
\newblock {Large g-factor enhancement in high-mobility InSb\textbackslash AlSb quantum wells}.
\newblock {\em Applied Physics Letters}, 81(10):1833--1835, 2002.

\bibitem{Aleshkin2008}
V.~Ya Aleshkin, V.~I. Gavrilenko, A.~V. Ikonnikov, S.~S. Krishtopenko, Yu~G. Sadofyev, and K.~E. Spirin.
\newblock {Exchange enhancement of the g factor in \ce{InAs}/\ce{AlSb} heterostructures}.
\newblock {\em Semiconductors}, 42(7):828--833, 2008.

\bibitem{Nizhankovskii1988}
V~I Nizhankovskii, B~K Medvedev, and V~G Mokerov.
\newblock Chemical-potential and g-factor of a 2d electron-gas in a strong magnetic field.
\newblock {\em JETP LETTERS}, 47(7):410--413, apr 1988.

\bibitem{Nizhankovskii2011}
V~I Nizhankovskii.
\newblock {Thermodynamics of Two-Dimensional Electron Gas in a Magnetic Field}.
\newblock {\em Physics Research International}, 2011:742158, 2011.

\bibitem{Nextnano}
Stefan Birner, Tobias Zibold, Till Andlauer, Tillmann Kubis, Matthias Sabathil, Alex Trellakis, and Peter Vogl.
\newblock nextnano: General purpose 3-d simulations.
\newblock {\em IEEE Transactions on Electron Devices}, 54(9):2137--2142, 2007.

\bibitem{Tersoff1985}
J.~Tersoff and D.~R. Hamann.
\newblock Theory of the scanning tunneling microscope.
\newblock {\em Phys. Rev. B}, 31:805--813, Jan 1985.

\bibitem{Becker2011}
S. Becker, C. Karrasch, T. Mashoff, M. Pratzer, M. Liebmann, V. Meden, M. Morgenstern
\newblock Probing Electron-Electron Interaction in Quantum Hall Systems with Scanning Tunneling Spectroscopy
\newblock {\em Phys. Rev. Lett.}, 106(15):156805--156808, 2011.

\bibitem{Bychkov1984}
Yu~A Bychkov and E~I Rashba.
\newblock {Oscillatory effects and the magnetic susceptibility of carriers in inversion layers}.
\newblock {\em Journal of Physics C: Solid State Physics}, 17(33):6039--6045, nov 1984.

\bibitem{Hernangomez-Perez2013}
Daniel Hernang{\'{o}}mez-P{\'{e}}rez, Jascha Ulrich, Serge Florens, and Thierry Champel.
\newblock {Spectral properties and local density of states of disordered quantum Hall systems with Rashba spin-orbit coupling}.
\newblock {\em Physical Review B}, 88(24):245433, dec 2013.

\bibitem{Ihn2009}
Thomas Ihn.
\newblock {\em {Semiconductor Nanostructures: Quantum states and electronic transport}}.
\newblock Oxford University Press, Oxford, 2009.

\bibitem{Hermann1977}
Claudine Hermann and Claude Weisbuch.
\newblock {\textbf{k.p} perturbation theory in III-V compounds and alloys: A reexamination}.
\newblock {\em Physical Review B}, 15(2):823--833, 1977.

\bibitem{Winkler2003}
Roland Winkler.
\newblock {\em Spin-orbit Coupling Effects in Two-Dimensional Electron and Hole Systems}.
\newblock Springer Berlin, Heidelberg, 2003.

\bibitem{Andrada1994}
E.~A. de~Andrada~e Silva, G.~C. La~Rocca, and F.~Bassani.
\newblock Spin-split subbands and magneto-oscillations in iii-v asymmetric heterostructures.
\newblock {\em Phys. Rev. B}, 50:8523--8533, Sep 1994.

\bibitem{Li2012}
Juerong Li, A.~M. Gilbertson, K.~L. Litvinenko, L.~F. Cohen, and S.~K. Clowes.
\newblock {Transverse focusing of spin-polarized photocurrents}.
\newblock {\em Physical Review B - Condensed Matter and Materials Physics}, 85(4):1--7, 2012.




\end{thebibliography}

\end{document}